\documentclass[aps, prl, twocolumn, showpacs, floatfix, superscriptaddress, groupedaddress]{revtex4-1}
\usepackage[utf8]{inputenc}
\usepackage{amsthm}
\usepackage{amssymb}
\usepackage{amsmath}
\usepackage{gensymb}
\usepackage{graphicx}
\usepackage{url}
\usepackage{framed}
\usepackage{float}
\usepackage{wrapfig}
\usepackage{multirow} 
\usepackage{tikz}
\usepackage{fullpage}
\usepackage[toc,page]{appendix}
\usepackage{physics}
\usepackage{listings}
\usepackage{hyperref}
\hypersetup{
    colorlinks=true,
    linkcolor=blue,
    citecolor=blue,
    filecolor=magenta,      
    urlcolor=blue,
}

\usepackage{natbib}

\usetikzlibrary{arrows}
\usetikzlibrary{decorations.markings}

\usepackage{csquotes}
\MakeOuterQuote{"}

\begin{document}


\title{Quantum Hall nano-interferometer in graphene}

\author{N. Moreau$^1$, B. Brun$^{1}$, S. Somanchi$^2$, K. Watanabe$^3$, T. Taniguchi$^4$, C. Stampfer$^2$ \& B. Hackens$^1$}

\affiliation{
$^1$ IMCN/NAPS, Universit\'e catholique de Louvain (UCLouvain), B-1348 Louvain-la-Neuve, Belgium \\
$^2$ JARA-FIT and 2nd Institute of Physics - RWTH Aachen, Germany \\
$^3$ Research Center for Functional Materials, National Institute for Materials Science, Namiki, Japan \\
$^4$ International Center for Materials Nanoarchitectonics, National Institute for Materials Science, Namiki, Japan
}

%

\date{\today}

\begin{abstract}
Quantum Hall edge states offer avenues for quasiparticle interferometry, provided that the ratio between phase coherence length and quantum Hall interferometer (QHI) size is large enough. Maximizing this ratio by shrinking the QHI area favors Coulomb interactions, impairing clear interferences observation. Here, we use scanning gate spectroscopy to probe interference regime in antidots-based graphene  nano-QHIs, free of localized states (LS). A simple Fabry-Perot model, without Coulomb interaction, reproduces the QHI phenomenology, even in the smallest QHI, highlighting the LS detrimental role.
\end{abstract}

\maketitle


In two-dimensional electronic systems (2DES), quantum Hall interferometers (QHIs) are the counterpart of optical interferometers, with coherent one-dimensional quantum Hall edge channels (QHECs) playing the role of monochromatic light beams. In QHIs, the beam-splitters are realized with quantum point contacts (QPCs) which bring the interfering QHECs in close proximity so as to modulate the reflection $R$ and the transmission $T=1-R$ of charge carriers in and out the interferometer \cite{Halperin2011}. In this framework, Fabry-Perot (FP) \cite{Wees1989} and Mach-Zehnder \cite{Ji2003} QHIs have been successfully implemented, leading to the observation of clear interferences due to Aharonov-Bohm (AB) effect. These breakthroughs have offered a promising path towards quantum computing based on anyonic braiding \cite{Baeuerle2018,Nakamura2020,Bartolomei2020}. Nevertheless, achieving this goal requires to meet conflicting specifications: a large coherence length, best reached in small QHIs \cite{Roulleau2008}, and no Coulomb charging effect, favored in large QHIs \cite{Zhang2009}. 

Recently, strategies have been elaborated to circumvent this apparent antagonism, and limit the effect of Coulomb charging. Electrically conductive planes have been added close to the semiconductor-based 2DES to screen the Coulomb effect of localized states in the bulk and lower their coupling with QHECs. With such architecture, the QHI minimal size at which the Coulomb dominated (CD) regime supplants AB effect has been substantially decreased \cite{Nakamura2019} (typically down to a $\mu$m lateral size). In the same vein, graphene devices using graphite back-gates provide similar advantages \cite{Deprez2021,Ronen2021}. Furthermore, the extra valley degeneracy and the possibility to mix electron and hole-filled QHECs with pn junction in this material let envision unprecedented realizations \cite{Williams2007,Oezyilmaz2007,Amet2014,Matsuo2015,Wei2017,Jo2021,Bours2017}. However, graphene edges turn out to be detrimental for robustness of topological protection of QHECs \cite{Polshyn2018,Zeng2019} and can impair the visibility of AB interference \cite{Ronen2021}. To explain this adverse effect, local probe measurements pointed towards the presence of both up- and downstream QHECs along the same edge of graphene devices, which weakens the topological protection of the downstream QHECs \cite{Cui2016,Marguerite2019}. More precisely, we have demonstrated that antidots trigger the coupling of these counterflowing QHECs for holes \cite{Moreau2021,Moreau2021a}.

In this letter, we use scanning gate microscopy (SGM) to show that these antidots can also form the basis of nano-FP QHIs, with a radius $r\sim$100 nm compatible with the spacing between up- and downstream QHECs and about one order of magnitude smaller in area than any QHI examined up to now. The extracted local source-drain transport spectroscopies indicate that such QHIs operate in the AB regime and their transport properties are ruled by the coupling intensity, i.e., the reflection $R$, between the antidot and the QHECs. We accurately reproduce the experimental data with a simple FP model which yields the QHEC velocity, found similar to the outcome of other studies. Observing the AB regime in $<100$ nm radius QHIs is surprising since CD regime should dominate in this case. We claim that the absence of localized state inside the antidot-based QHI drastically reduces the Coulomb interaction. These results let envision the design of nano-sized QHIs in which coherent oscillations exhibit high visibility without being impaired by Coulomb interactions.

\begin{figure}[!ht]
\centering
\includegraphics[width=\linewidth]{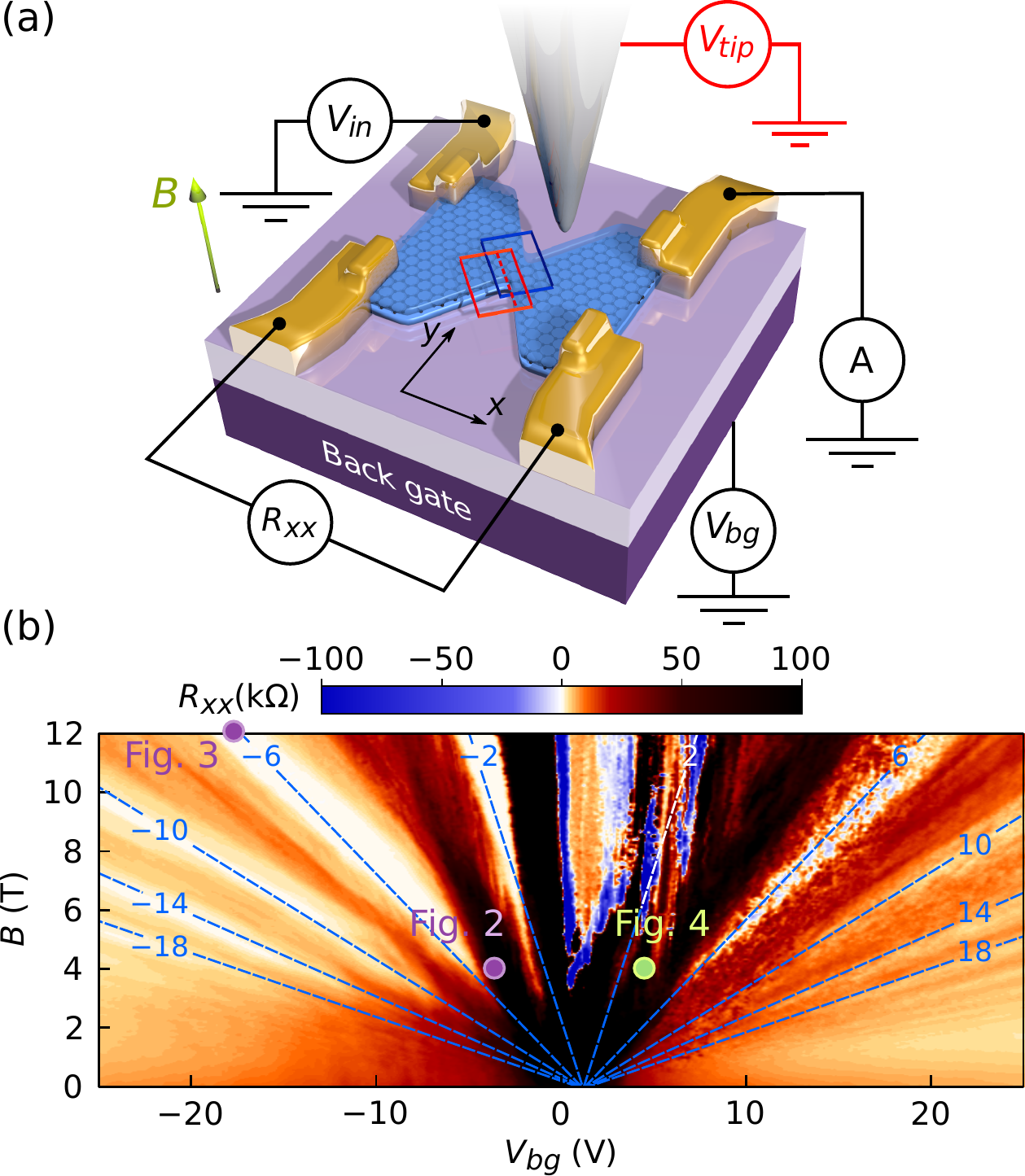}
\caption{(a) Artist view of the experimental setup. (b) $R_{xx}$ map as a function of $V_{bg}$ and $B$. Blue lines indicate the filling factors $\nu = \pm (4n+2)$ obtained by a fit of the $R_{xx} = 0$ regions.}
\label{sample}
\end{figure}

The studied sample, depicted in Figs. \ref{sample}a,b, consists of a monolayer of graphene encapsulated between two hBN flakes \cite{Terres2016}. A constriction geometry has been lithographically defined but has no influence for hole charge carriers \cite{Moreau2021a}. The longitudinal resistance $R_{xx}$ is measured via line contacts \cite{Wang2013}. Global charge carrier density is varied using a back gate voltage $V_{bg}$ and a sharp metallic SGM tip, biased at a voltage $V_{tip}$, is used to change the local charge carrier density in the vicinity of the tip (Fig. \ref{sample}a). A magnetic field $B$ is applied perpendicularly to the graphene plane and leads to the fan diagram shown in Fig. \ref{sample}b, discussed in details in ref. \cite{Moreau2021a}. 


\begin{figure}[!ht]
\centering
\includegraphics[width=\linewidth]{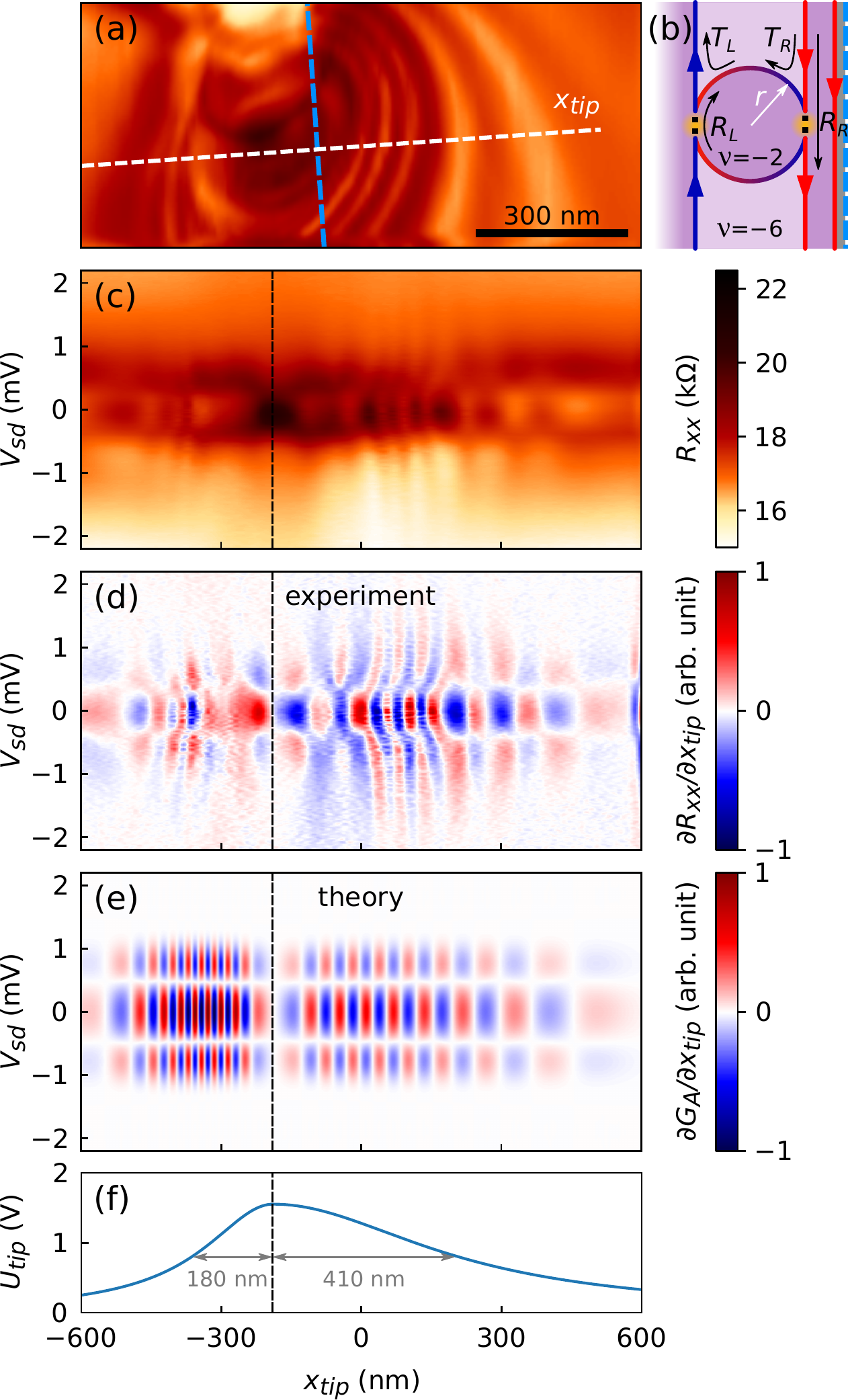}
\caption{(a) SGM maps of $R_{xx}$ as a function of tip position, with $B = 4$ T, $V_{bg} = -3.6$V (see Fig. \ref{sample}b) and $V_{tip} = 0$ V. The scan area is sketched by a red rectangle (limited to the dashed line) in Fig. \ref{sample}b. (b) Schematic of the antidot located between the upstream (blue) and downstream (red) QHECs. It forms a FP QHI strongly coupled to the QHECs, i.e., with low reflection coefficients at the two beam splitter depicted with black doted lines highlighted in yellow. (c) $R_{xx}$ as a function of $V_{sd}$ and tip position $x_{tip}$ along the white dashed line in (a). (d) Derivative of $R_{xx}$ in (c) according to $x_{tip}$. (e) Theoretical map of $\partial G_{A}/\partial x_{tip}$ as a function of $V_{sd}$ and $x_{tip}$. (f) Tip potential ($U_{tip}$ as a function of $x_{tip}$) used in (e).}
\label{checkerboard}
\end{figure}

First, we use SGM to study hole transmission from down- to upstream QHECs at $B = 4$ T. Fig. \ref{checkerboard}a shows a $R_{xx}$ map obtained by scanning the tip above one of the device edges. The concentric circular fringes are the signatures of transport through an antidot coupling the up- and downstream QHECs, as depicted in Fig. \ref{checkerboard}b \cite{Hackens2010,Martins2013,Moreau2021,Moreau2021a}. In this configuration, we will now show that the antidot behaves as a FP QHI where the left and right tunneling paths (black dotted lines highlighted in yellow in Fig. \ref{checkerboard}b) act as beam splitters. An in-depth characterization of transport in this system is achieved via a local (tip-controlled) source-drain spectroscopy, shown in Fig. \ref{checkerboard}c. It has been obtained by varying the DC source-drain bias voltage $V_{sd}$ while moving the tip along the white dashed line in Figs. \ref{checkerboard}a, i.e., crossing the location of the antidot, at constant $V_{tip}$. The derivative of the map in Fig. \ref{checkerboard}c with respect to $x_{tip}$  (Fig. \ref{checkerboard}d) exhibits the typical checkerboard-like pattern observed in QHI in the AB regime \cite{Neder2006,Roulleau2007,McClure2009,Martins2013,Nakamura2019,Jo2021,Deprez2021,Ronen2021}. Figure \ref{checkerboard}d features AB oscillations vs $x_{tip}$, caused by a variation of the magnetic flux $\Phi = BA$ enclosed in the antidot of area $A$, as well as oscillations vs $V_{sd}$, linked to a bias-driven phase shift of $L\epsilon /\hbar v$ where $L$ is the length of one of the QHI path, $\epsilon$ is the energy and $v$ the QHEC velocity \cite{Moreau2021b}.

Now, we demonstrate that numerical simulations of the conductance $G_A$ through the antidot, based on a simple FP model further detailed in ref. \cite{Moreau2021b}, accurately reproduce the experimental data, as shown in Fig. \ref{checkerboard}e. This map is obtained with small reflection coefficients at the left and right beam splitters ($R_L = R_R = 0.01$ - see Fig. \ref{checkerboard}b) and with an antidot radius $r = 100$ nm, in full accordance with the expected distance between up- and downstream QHECs estimated between 200 and 300 nm \cite{Cui2016, Marguerite2019, AharonSteinberg2021, Moreau2021}. The oscillation period vs $x_{tip}$ have been fitted by taking into account a change of the antidot radius (and hence its area $A$) caused by the tip-induced electrostatic potential $U_{tip}$ ($r$ changes from 100 nm to 108 nm at the maximum value of $U_{tip}$). The conversion from $x_{tip}$ to $U_{tip}$ axis has been performed using the tip potential profile shown in Fig. \ref{checkerboard}f (see supplemental materials). Finally, the oscillations vs $V_{sd}$ have been fitted with a QHEC velocity $v = 1.5\cdot 10^5$ m/s, in good agreement with former experimental results in semiconductor-based heterostructures \cite{McClure2009,Martins2013,Gurman2016,Nakamura2019} and in graphene \cite{Deprez2021,Ronen2021}. Note that there is a direct correspondance between $R_{xx}$ and $G_A$ since a large conductance through the antidot yields backscattering from the down- to the upstream QHEC and, hence, a large $R_{xx}$.

Another interesting feature of Fig. \ref{checkerboard}d is the decrease of the oscillations visibility when $|V_{sd}|$ is raised. The precise decoherence mechanism leading to this behavior is debated in the literature \cite{Youn2008,Levkivskyi2008,Kovrizhin2009,Schneider2011} and is usually captured in numerical calculations by multiplying the pattern obtained with the FP model by $\exp(-\gamma |V_{sd}|^n)$, with $n=1$ or 2 \cite{McClure2009,Martins2013,Huynh2012}. Here, we have chosen another approach, inspired by ref. \cite{Roulleau2007}, consisting in introducing a gaussian dephasing whose variance is proportional to $V_{sd}$ (see supplemental materials). This approach has provided better results with the simulations presented below.  

\begin{figure}[!ht]
\centering
\includegraphics[width=\linewidth]{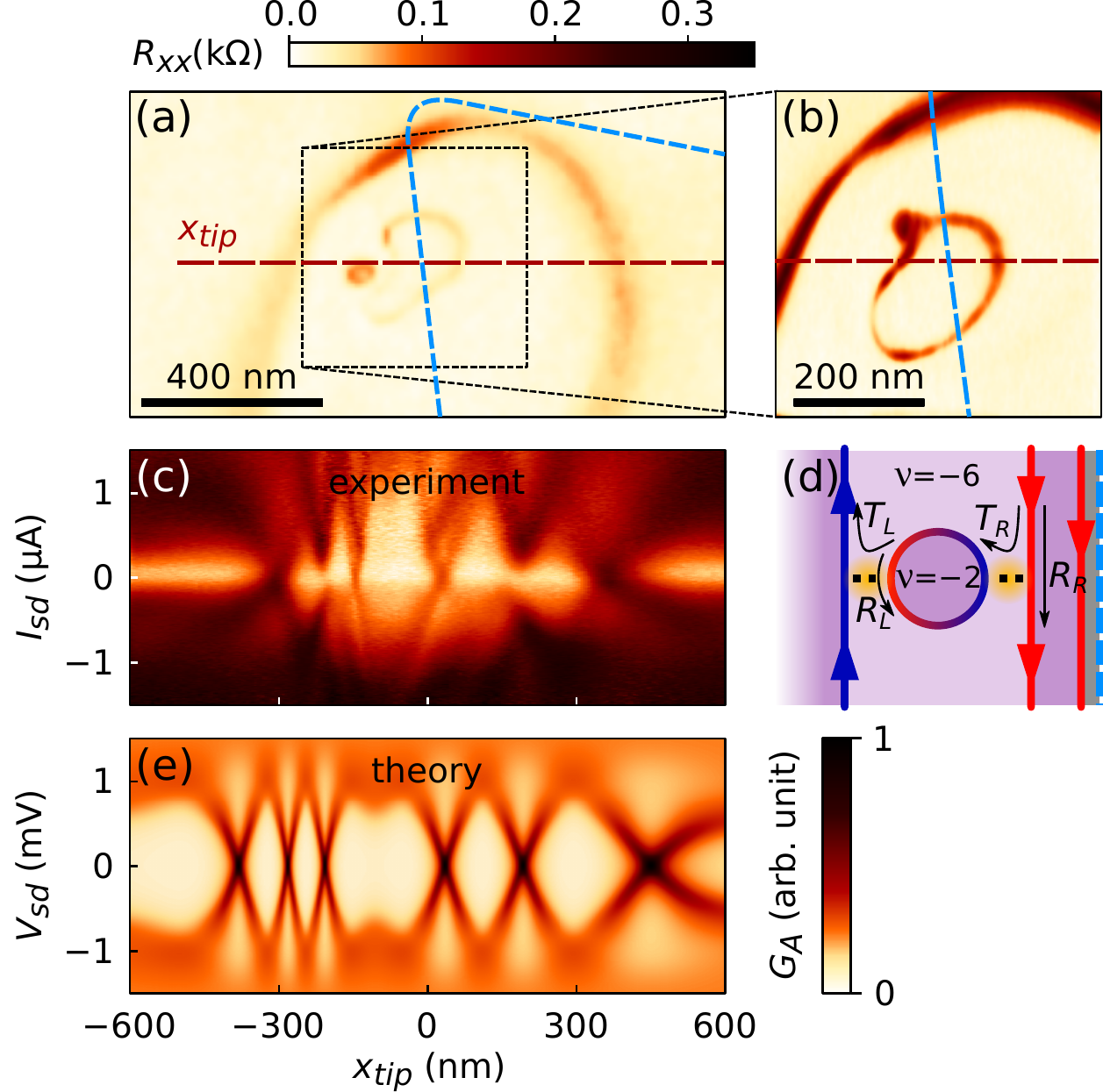}
\caption{(a) SGM maps of $R_{xx}$ as a function of tip position, with $B = 12$ T, $V_{bg} = -17.75$V (see Fig. \ref{sample}b) and $V_{tip} = 0$ V. The scan area is sketched by a red rectangle in Fig. \ref{sample}b. (b) zoom in (a). (c) $R_{xx}$ as a function of $I_{sd}$ and tip position $x_{tip}$ along the red dashed line in (a). Note that, due to the low $R_{xx}$, the spectroscopy is performed by applying a DC current $I_{sd}$ instead of a voltage bias. (d) Schematic of the FP QHI weekly coupled to the QHECs, i.e., with large reflection coefficients at the two beam splitter depicted with black doted lines highlighted in yellow. (e) Theoretical map of $G_{A}$ as a function of $V_{sd}$ and $x_{tip}$.}
\label{diamonds}
\end{figure}

In the following, we discuss the influence of the reflection coefficients $R_L$ and $R_R$ of the beam splitters, considering the experimental data obtained at different values of $B$ and $V_{bg}$. First, Fig. \ref{diamonds} is obtained in similar conditions as Fig. \ref{checkerboard} but at larger magnetic field ($B = 12$ T). The SGM maps in Figs. \ref{diamonds}a,b feature a $R_{xx} = 0$ background with sharp concentric non-zero $R_{xx}$ fringes centered on the antidot position. The local spectroscopy map shown in Fig. \ref{diamonds}c exhibits a clear diamonds pattern, in sharp contrast with the checkerboard pattern of Fig. \ref{checkerboard}. This kind of pattern is usually associated to Coulomb charging effects \cite{Kataoka1999,McClure2009,Martins2013a}. However, we show in ref. \cite{Moreau2021b} that the numerical model mentioned above, considering a pure AB regime, also reveal diamonds features for large reflection coefficients at the beam splitters of the QHI. It is confirmed by the map shown in Fig. \ref{diamonds}e calculated with the same equation as the one used to generate Fig. \ref{checkerboard}e, but with $R_L = R_R = 0.7$. Here, the antidot radius increases from $r = 57.7$ nm to $59.5$ nm with the change of tip-induced potential following the same evolution in $x_{tip}$ as depicted in Fig. \ref{checkerboard}f. Note that for the purpose of the simulation, the vertical scale of Fig. \ref{diamonds}e should be expressed in $V_{sd}$ whereas the map in Fig. \ref{diamonds}c has been obtained by varying $I_{sd}$ due to the presence of $R_{xx} = 0$ regions. Therefore, the QHECs velocity found in this case ($v = 1.4\cdot 10^{5}$ m/s) suffers from a large uncertainty.  

At this stage, it remains to explain what is the rationale behind the change of reflection coefficients $R_L$ and $R_R$ between Fig. \ref{checkerboard} and \ref{diamonds}. It can originate from two factors at least. First, the size of the antidot and the distance between the up- and downstream QHECs depend on the position of the Landau levels compared to the Fermi energy \cite{Moreau2021b}. The proximity between the antidot and the QHECs, and hence the value of $R_L$ and $R_R$, is consequently different for each couple of $B$ and $V_{bg}$. Second, the width of the QHECs wavefunction decreases at larger magnetic field, since it depends on the magnetic length $l_B\propto B^{-1/2}$. Therefore, the overlap between the antidot and the QHECs wavefunctions also decreases with $B$ (by a factor $\sqrt{3}$ between $B = 4$ and 12 T), leading to an increase of the coefficients $R_L$ and $R_R$ (the transmission from the QHEC to the antidot drops), in agreement with the strong versus low coupling regimes observed between Fig. \ref{checkerboard} and \ref{diamonds} respectively.

\begin{figure}[!ht]
\centering
\includegraphics[width=\linewidth]{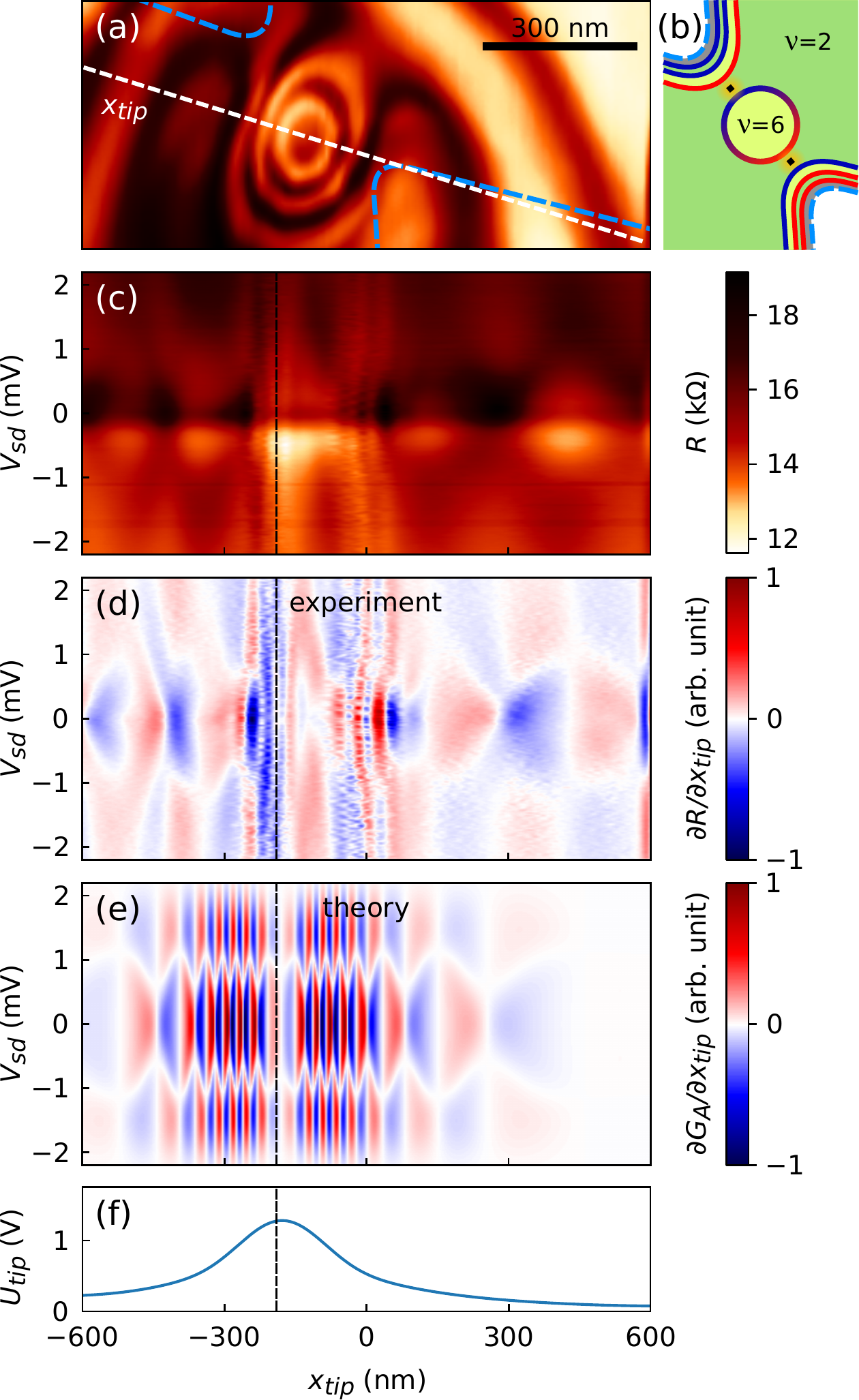}
\caption{(a) SGM maps of $R_{xx}$ as a function of tip position, with $B = 4$ T, $V_{bg} = 4.5$V (see Fig. \ref{sample}b) and $V_{tip} = 0$ V. The scan area is sketched by a blue rectangle in Fig. \ref{sample}b. (b) Schematic of the antidot located in the center of the constriction, between the upstream QHECs of both edges. It forms a FP QHI with intermediate reflection coefficients at the two beam splitter. (c) $R_{xx}$ as a function of $V_{sd}$ and tip position $x_{tip}$ along the white dashed line in (a). (d) Derivative of $R_{xx}$ in (c) according to $x_{tip}$. (e) Theoretical map of $\partial G_{A}/\partial x_{tip}$ as a function of $V_{sd}$ and $x_{tip}$. (f) Tip potential ($U_{tip}$ as a function of $x_{tip}$) used in (e).}
\label{mix}
\end{figure}

Next, we examine the electron side in Fig. \ref{mix} at $B = 4$ T. The SGM map in Fig. \ref{mix}a, obtained at positive $V_{bg}$, shows fringes centered in the vicinity of the constriction, in opposition to the contrast located along the edges in the hole side (Figs. \ref{checkerboard}a and \ref{diamonds}a,b). In ref. \cite{Moreau2021a}, we ascribe this difference to an asymmetry in the QHEC structure between the electron and hole sides caused by the electrical contact. In that framework, the electron backscattering is mediated by an antidot located in the vicinity of the constriction (Fig. \ref{mix}b) that couples the upstream QHECs transmitted along the etched opposite edges of the QPCs. The spectroscopy map meadured on this antidot is shown in Fig. \ref{mix}c and its derivative vs $x_{tip}$ in Fig. \ref{mix}d. The latter map features a strained checkerboard pattern well captured by the numerical simulation of Fig. \ref{mix}e. It has been obtained with the intermediate reflection coefficients $R_L = R_R = 0.15$ and with an antidot radius going from $r = 70$ nm to $80.45$ nm under the action of the tip-induced potential depicted in Fig. \ref{mix}g. The QHECs velocity, extracted from a fit to the experimental oscillations vs $V_{sd}$, is $v = 2\cdot 10^{5}$ m/s, i.e., similar to values found on the hole side.

Nevertheless, finding the studied antidots in the AB regime in the different configurations discussed in Figs. \ref{checkerboard}, \ref{diamonds} and \ref{mix} is unexpected in regards to their small sizes ($A < 0.05 ~ \mu \text{m}^2$). Indeed, up to our knowledge, no QHI operating in this regime has ever been reported with a size under $1 \mu \text{m}^2$, even with architectures optimized to reduce Coulomb interactions \cite{Nakamura2019,Deprez2021,Ronen2021}. To better understand the transition between both regimes, one can consider the relationship \cite{Halperin2011}
\begin{equation}\label{xi}
    \xi = \dfrac{C_{eb}}{C_{eb}+C_b},
\end{equation}
where $C_{eb}$ is the capacitive coupling between the QHECs and the localized bulk states, which is proportional to the QHI perimeter (here, $C_{eb}\propto r$), and $C_b$ is the bulk capacitance, proportional to the QHI area (here, $C_{b}\propto r^2$). The QHI operates in the AB (CD) regime when $\xi \rightarrow 0$ ($\xi \rightarrow 1$). By decreasing the QHI size, the coefficient $C_{b}$ becomes negligible compared to $C_{eb}$ in Eq. \ref{xi} and the CD regime dominates. However, Eq. \ref{xi} is only valid provided that localized states are present inside the QHI. In the case of an antidot, this is mostly unlikely since its size is similar with the disorder potential fluctuations observed in graphene on hBN \cite{Xue2011,Moreau2021}. In other words, the QHI is actually made of a localized state. Hence, $C_{eb} = \xi = 0$ in Eq. \ref{xi}, which supports the observation of the AB regime in our nano-size QHI.

In summary, we have used SGM to reveal the presence of preexisting nano-QHI, associated with antidots, causing the coupling between up- and downstream QHECs in graphene. From the local spectroscopies performed on these QHI, accurately reproduced with a simple FP numerical model, we find that even the smallest-radius (sub-100 nm) QHI operates in the AB regime. From the fit to experimental data, we have extracted both a value for the QHEC velocity, similar to previously extracted values (in the case of graphene QHI), and for the antidot size, coherent with the expected distance between up- and downstream QHECs but much smaller than previously investigated QHI in the AB regime. According to leading theories in the field, QHI of such small size should lead to a CD regime. We explain this apparent contradiction by the absence of localized state inside the antidot, whose length scale is on the order of the disorder potential fluctuations. 

Based on these results, new designs of QHIs operating in the AB regime can be envisioned in graphene, with unprecedented small sizes offering no compromise between high visibility and the absence of Coulomb effect. The up- and downstream modes appearing spontaneously at graphene edges can be replaced by counterpropagating QHECs controlled by top gates \cite{Zimmermann2017} and the antidot-based nano-QHI can be easily introduced by injecting charges into the hBN layer, as recently demonstrated in refs. \cite{Gutierrez2018,Walkup2020}. It opens the way to the realization of complex systems where several QHIs can interact with each other while conserving the phase coherence thanks to the small size of such architecture.

\begin{acknowledgments}
This work was partly funded by the Federation Wallonie-Bruxelles through the ARC Grant No. 16/21-077, by the F.R.S-FNRS through the Grant No. J008019F, and from the European Union’s Horizon 2020 Research and Innovation program (Core 1 No. 696656 and Core 2 No. 785219). This work was also partly supported by the FLAG-ERA grant TATTOOS, through F.R.S.-FNRS PINT-MULTI grant No. R 8010.19. B.B. (research assistant), B.H. (research associate) and N.M. (FRIA fellowship) acknowledge financial support from the F.R.S.-FNRS of Belgium. Support by the Helmholtz Nanoelectronic Facility (HNF), the EU ITN SPINOGRAPH and the DFG (SPP-1459) is gratefully acknowledged. K.W. and T.T. acknowledge support from the Elemental Strategy Initiative conducted by the MEXT,
Japan ,Grant Number JPMXP0112101001, JSPS KAKENHI Grant Numbers JP20H00354 and the CREST(JPMJCR15F3), JST.

\end{acknowledgments}

\end{document}


\title{\textbf{\LARGE{Quantum Hall nano-interferometer in graphene \\ - Supplementary materials -}}}

\author{\large{N. Moreau$^1$, B. Brun$^{1}$, S. Somanchi$^2$, K. Watanabe$^3$, }\\ \large{T. Taniguchi$^4$, C. Stampfer$^2$ \& B. Hackens$^1$} \\
\small{$^1$IMCN/NAPS, Universit\'e catholique de Louvain (UCLouvain), B-1348 Louvain-la-Neuve, Belgium} \\
\small{$^2$ JARA-FIT and 2nd Institute of Physics - RWTH Aachen, Germany} \\
\small{$^3$ Research Center for Functional Materials,}\\
\small{National Institute for Materials Science, 1-1 Namiki, Japan} \\
\small{$^4$ International Center for Materials Nanoarchitectonics,}\\
\small{National Institute for Materials Science, 1-1 Namiki, Japan}}


\date{}

\maketitle

\section{Experimental setup}
The studied sample consists in a monolayer of graphene, encapsulated between two 20 nm-thick hBN flakes, in which a 250 nm-wide constriction geometry has been lithographically-defined \cite{Terres2016}, as depicted in Fig. 1a of the manuscript. It is anchored to the mixing chamber of a dilution fridge whose base temperature is $\sim 100~\text{mK}$. A magnetic field $B$ is applied perpendicularly to the graphene plane. Current is injected and voltage is measured via line contacts made of gold deposited on a thin adhesion layer of chromium \cite{Wang2013,Moreau2021a}. The four contacts allow to measure the longitudinal resistance $R_{xx}$ using the standard lock-in technique at $77~\text{Hz}$. The spectroscopies are obtained by adding a DC current on top the AC excitation. We then record the DC voltage drop between the two measurement contacts (excepted for Fig. 3 of the manuscript where the voltage has not been recorded). The global charge carrier density in graphene is varied by applying a voltage $V_{bg}$ on the silicon back-gate separated from the sample by a 300 nm-thick SiO$_2$ layer. We furthermore use a sharp metallic SGM tip, biased at a voltage $V_{tip}$, and scanning at a distance of 70 nm from the graphene plane, to change locally the charge carrier density.

\section{The tip-induced potential}\label{tippot}
The tip-induced potentials used in Figs. 1-3 of the manuscript have been obtained from Lorentzian functions of different half-width at half-maximum (HWHM) to capture the change of screening depending if the tip is above or away from the graphene plane. In Figs. 2e,f, 3e and 4e,f of the manuscript, the potential has been obtained with two half-Lorentzian as
\begin{equation}\label{Vtip}
    V^*_{tip}(x_{tip}) = 
    \begin{cases} 
    \dfrac{V_{max}}{1+((x_{tip}-x_A)/R_{tip,1})^2}, & {\rm for} ~ x_{tip}<x_c, \\
    \dfrac{V_{max,2}}{1+((x_{tip}-x_A)/R_{tip,2})^2}, & {\rm for} ~ x_{tip}>x_c, \\
    \end{cases}
\end{equation}
where $V_{max}$ is the maximal tip-induced potential ($V_{max}=-1.55$ V in all the maps), $x_A$ is the antidot position ($x_A = -230$ nm in Figs. 2e,f, $x_A = -150$ nm in Fig. 3e and $x_A = -180$ nm in Fig. 4e,f), $x_c$ is the edge of the graphene plane (we have taken $x_c = x_A$ for Figs. 2e,f and 3e and $x_c = 20$ nm for Figs. 4e,f), $R_{tip,1}$ is the HWHM above the graphene plane ($R_{tip,1} = 125$ nm in all the maps), $R_{tip,2}$ is the HWHM outside the graphene plane ($R_{tip,2} = 350$ nm in Figs. 2e,f and 3e and $R_{tip,2} = 250$ nm since the tip is moved along the edge of the device - see Fig. 4a) and $V_{max,2}$ is the amplitude of the Lorentzian function outside the graphene plane that guarantees the continuity of the function at $x = x_c$ ($V_{max,2} = V_{max}$ in Figs. 2e,f and 3e and $V_{max,2} = -0.714$ V. 

However, Eq. \ref{Vtip} does not guarantee the continuity of $\partial V_{tip}/\partial x_{tip}$, since it considers an abrupt transition between the graphene plane where screening is at play and away from the graphene plane (or along its edge for Fig. 4). To overcome this issue, we perform a convolution of $V_{tip}(x_{tip})$ with a Gaussian, giving
\begin{equation}
    V_{tip}(x_{tip}) = \frac{\displaystyle {\int_{-\infty}^\infty V^*_{tip}(x)\exp\left(-(x-x_{tip})^2/\Delta x^2\right)dx}}{{\displaystyle \int_{-\infty}^\infty \exp\left(-x^2/\Delta x^2\right)dx}}
\end{equation}
where $\Delta x$ is the width of the smoothing function. In all the maps of the manuscript, we have taken $\Delta x 45$ nm.

\section{The Fabry-Perot model}
The model used to obtain the maps in the manuscript is detailed in ref. \textbf{add ref}. It allows to compute the conductance $G$ through a FP quantum Hall intereformeter (QHI) as a function of the DC bias $V_{sd}$, and yields
\begin{equation}
    G = \frac{-e}{h} \frac{d}{dV_{sd}} {\displaystyle\bigintss_{-eV_{sd}/2}^{eV_{sd}/2}\dfrac{(1-R_L)(1-R_R)}{1+R_LR_R-2\sqrt{R_LR_R}\cos\left(\pi\frac{\Phi}{\Phi_0}+\frac{L\epsilon}{\hbar v} + \varphi\right)}- \frac{(1 - R_L)(1 - R_R)}{1 + R_LR_R} d\epsilon},
\end{equation}
where $e$ is the charge of the electron, $h$ the Plank constant, $R_L$ ($R_R$) the reflection probability at the left (right) QHI beam splitter, $\Phi$ is the magnetic flux enclosed in the QHI, $\phi_0 = h/2e$ is the quantum of magnetic flux, $L$ the length of one of the QHI arm, $v$ the charge carriers velocity and $\varphi$ a phase. 

Importantly, the magnetic flux $\Phi = BA$, where $A \approx L^2/\pi$ is the QHI area (by considering a circular geometry) is modified by the tip-induced potential determined in section \ref{tippot}. As justified in more details in ref \textbf{add ref}, we consider a linear evolution of the radius $R = L/\pi$ with $V_{tip}$ and $B$ and we obtain
\begin{equation}\label{LVtipB}
    R = R_0 + \alpha_{tip} \Delta V_{tip} + \alpha_B \Delta B.
\end{equation}
where $R_0$ is the initial QHI radius at $V_{tip} = 0$ and $\alpha_{tip}$ and $\alpha_B$ are proportional coefficients. $\alpha_B$ is inconsistent since $B$ remains constant in each maps and $\alpha_{tip}$ is obtained by determining a maximum value of $R$ when $V_{tip} = V_{max}$ (see Eq. \ref{Vtip}). For each map, we have used the following parameters that are also given in the manuscript
\begin{table}[!ht]
\centering
\begin{tabular}{lll}
\hline
 & $V_{tip} = 0$ & $V_{tip} = V_{max}$ \\
\hline
Fig. 2e & $R_0 = 100$ nm & $R = 108$ nm \\
Fig. 3e & $R_0 = 57.7$ nm & $R = 59.5$ nm \\
Fig. 4e & $R_0 = 70$ nm & $R = 80.5$ nm \\
\hline
\end{tabular}
\end{table}

\section{The vanishing visibility}
As explained in the manuscript, we have captured the decrease of oscillation visibility in $V_{sd}$ by applying a Gaussian averaging whose variance is proportional to $V_{sd}^2$. By departing from the map $G^*(V_{sd},x_{tip})$ obtained from the Fabry-Perot model developed in \textbf{add ref}, it yields
\begin{equation}
    G(V_{sd},x_{tip}) = \frac{\displaystyle {\bigintss_{-\infty}^\infty G^*(\epsilon,x_{tip})\exp\left(-\left(\frac{\epsilon - V_{sd}}{\gamma V_{sd}^2}\right)^2\right)d\epsilon}}{{\displaystyle \bigintss_{-\infty}^\infty \exp\left(-\left(\frac{\epsilon - V_{sd}}{\gamma V_{sd}^2}\right)^2\right)d\epsilon}}
\end{equation}
where $\gamma$ is a fitting parameter. Figs. 2e and 3e of the manuscript have been obtained with $\gamma = 500$ whereas Fig. 4e has been obtained with $\gamma = 250$.